\documentclass[10pt]{article}
\usepackage{dcolumn}
\usepackage{bm}
\usepackage{verbatim}       

\usepackage[margin=4.3cm]{geometry}

\usepackage[dvips]{graphicx}

\usepackage{amsthm}

\usepackage{bm}
\usepackage{amstext}
\usepackage{psfrag}
\usepackage{amsmath}
\usepackage{amssymb}
\usepackage{amsfonts}

\newcommand{\dd}{{\rm d}}

\newcommand{\bd}{\begin{definition}}                
\newcommand{\ed}{\end{definition}}                  
\newcommand{\bc}{\begin{corollary}}                 
\newcommand{\ec}{\end{corollary}}                   
\newcommand{\bl}{\begin{lemma}}                     
\newcommand{\el}{\end{lemma}}                       
\newcommand{\bp}{\begin{proposition}}            
\newcommand{\ep}{\end{proposition}}                
\newcommand{\bere}{\begin{remark}}                  
\newcommand{\ere}{\end{remark}}                     

\newcommand{\bt}{\begin{theorem}}
\newcommand{\et}{\end{theorem}}

\newcommand{\be}{\begin{equation}}
\newcommand{\ee}{\end{equation}}

\newcommand{\bit}{\begin{itemize}}
\newcommand{\eit}{\end{itemize}}
\newtheorem{theorem}{Theorem}[section]
\newtheorem{corollary}[theorem]{Corollary}
\newtheorem{lemma}[theorem]{Lemma}
\newtheorem{proposition}[theorem]{Proposition}
\theoremstyle{definition}
\newtheorem{definition}[theorem]{Definition}
\theoremstyle{remark}
\newtheorem{remark}[theorem]{Remark}

\begin{document}

\title{Product posets and causal automorphisms of the plane}

\author{Alfonso Garc{\'\i}a-Parrado G\'omez-Lobo\thanks{Centro de Matem\'atica, Universidade do Minho, 4710-057 Braga,
Portugal. E-mail: alfonso@math.uminho.pt} \ and Ettore Minguzzi\thanks{
Dipartimento di Matematica Applicata ``G. Sansone'', Universit\`a
degli Studi di Firenze, Via S. Marta 3, I-50139 Firenze, Italy.
E-mail: ettore.minguzzi@unifi.it} }

\date{}

\maketitle

\begin{abstract}
\noindent 
A simple characterization of the causal automorphisms of 1+1
Minkowski spacetime is given.
\end{abstract}

\section{Introduction}

In this work by poset $P$ we mean a partially ordered set, that is a
set endowed with a reflexive, transitive and antisymmetric relation.
An order automorphism of $P$ is a surjective map $f: P \to P$ with
the property that $p\le q \Leftrightarrow f(p)\le f(q)$. An order
automorphism is necessarily injective. Indeed, $f(p)=f(q)$ implies
``$f(p)\le f(q)$ and $f(q)\le f(p)$'' from which we  deduce ``$p\le
q$ and $q\le p$'' and hence $p=q$. Given an order automorphism $f$
the inverse $f^{-1}$ is also an order automorphism.

A spacetime $(M,g)$ endowed with the causal order $\le$, i.e. $p\le
q$ if there is a future directed causal curve connecting $p$ to $q$
or $p=q$, is called causal if $(M,\le)$ is a poset. The order
automorphisms of a causal spacetime endowed with the causal relation
are then called causal automorphisms. We shall not demand that these
maps be continuous but  continuity will follow.

The causal automorphisms of Minkowski $n+1$ spacetime for $n\ge 2$
have been shown by Alexandrov
\cite{alexandrov50,alexandrov67,alexandrov67b} and Zeeman
\cite{zeeman64} to be generated by the inhomogeneous Lorentz group
and dilatations. Under the assumption of differentiability this
result was previously obtained by Liouville and Lie
\cite{dubrovin84} so that Alexandrov-Zeeman theorem follows from
Lioville-Lie theorem by noting that every bijective map between
strongly causal spacetimes which sends null geodesics into null
geodesics and conversely is smooth \cite{hawking76,huang98}.

Recently, Do-Hyung Kim has investigated the causal automorphism of
Minkowski 1+1 spacetime, namely $M=\mathbb{R}^2$  with coordinates
$(x^+,x^-)$ and metric $g=-\dd x^+\dd x^-$ (we use preferably light
cone coordinates $x^{\pm}=t\pm x$). With respect to the higher
dimensional case here we have the complication that the causal
automorphism need not be smooth. Kim reaches a simple and intuitive
result. Nevertheless, his proof is scattered over three papers
\cite{kim09,kim10,kim10b}, requires several propositions, and uses
 a tool of ``causally admissible system'' developed by him and connected to the representability of spacetime points
with compact sets on a Cauchy hypersurface.\footnote{Kim essentially
uses the light cone coordinates $x\pm t$ instead of $t\pm x$, as a
consequence the statement of his result needs to contemplate a case
in which certain homeomorphisms are decreasing, a complication that
does not appear in our formulation. Also note that he works on
$\mathbb{R}^2$ with coordinates $(x,t)$ not $(t,x)$.} Given the
simplicity of the final result it is natural to wonder whether it
could  be obtained through a simple direct proof. We provide such a
proof. In the end we obtain

\begin{theorem} \label{theo:main-result}
The causal automorphism of Minkowski 1+1 spacetime are generated by
the space reflection $(x^{+},x^-) \to (x^{-},x^+)$ and by the maps
$(x^+,x^-) \to (f(x^+),g(x^-))$ where $f,g:\mathbb{R}\to \mathbb{R}$
are increasing homeomorphisms of the real line.
\end{theorem}

Note that we do not demand that the causal automorphism be
continuous but the continuity follows. The fact that the space
reflection $(x^{+},x^-) \to (x^{-},x^+)$ is a causal automorphism
can be easily checked. Also the fact that the map $(x^+,x^-) \to
(f(x^+),g(x^-))$ is a causal automorphism whenever $f$ and $g$ are
increasing homeomorphisms of the real line can be easily verified.
The proof of the theorem is based on the observation that Minkowski
spacetime, as it is evident in light cone coordinates, is the
product poset of the real line (with the usual order) with itself.
Recall that if $A$ and $B$ are two posets then $A\times B$ can be
given the product order: $(a,b)\le (a',b')$ if $a\le a'$ and $b \le
b'$. The causal order has indeed this structure because
$(x^+,x^-)\le  (y^+,y^-)$ iff $x^+\le y^+$ and $x^-\le y^-$.

\begin{proof}
Let $F:\mathbb{R}^2\to \mathbb{R}^2$ be a causal automorphism (in
light cone coordinates). Let $O=(0,0)$ and define $G:\mathbb{R}^2
\to \mathbb{R}^2$ by $G(p)=F(p)-F(O)$ then since translations
preserve the causal relation we have that $G$ is a causal
automorphism that maps $O$ to itself. In a spacetime the horismos
relation is the set $E^{+}=J^{+}\backslash I^+$, in particular
 $(x,y)\in E^{+}$ if and only if there is a future
directed achronal lightlike geodesic connecting $x$ to $y$ (see
\cite{hawking73}). In Minkowski spacetime we have $(p,q)\in E^{+}$
if and only if $J^+(p)\cap J^{-}(q)$ is totally ordered by the
causal relation $\le$ (for generic causal spacetimes a similar
characterization of $E^+$ holds, see \cite{kronheimer67}
\cite[Theor. 3.9]{minguzzi06c}). Since $G$ and $G^{-1}$ preserve the
causal relation $\le (=J^+)$ and $E^{+}$ can be completely
characterized through it,  it follows that $(p,q)\in E^{+}$ if and
only if $(G(p),G(q))\in E^{+}$.

The axes passing through $O$ form the set $E:=E^{+}(O)\cup E^{-}(O)$
and since $O$ is mapped into itself we find that $E$ is mapped into
itself. Let us consider any two distinct points $p,q\ne O$ in one of
the axis passing through $O$. We can assume without loss of
generality that $J^{+}(p)\cap J^{-}(q)\ne \emptyset$ otherwise we
exchange  $p$ and $q$. Since $J^{+}(p)\cap J^{-}(q)$ is totally
ordered by $\le$ the images $G(p)$, $G(q)$, must satisfy
$J^{+}(G(p))\cap J^{-}(G(q))$ is non-empty and totally ordered by
$\le$ which is impossible if $G(p)$ and $G(q)$ do not lie on the
same axis. We conclude that each axis passing through $O$ is sent
into another axis passing though $O$ with a possible exchange of
axes. It is trivial that the space reflection $(x^{+},x^-) \to
(x^{-},x^+)$ is a causal automorphism and
 thus by applying it once if necessary, we can
always assume that the axes do not  get interchanged by $G$.

Now, consider the axis whose elements are given by $(x^+,0)$. This
axis is mapped by $G$ into itself,  thus there is a function
$u:\mathbb{R} \to \mathbb{R}$ such that $(x^+,0)\to (u(x^+),0)$.
Analogously, there is a function $v:\mathbb{R} \to \mathbb{R}$ such
that $(0,x^-)\to (0,v(x^-))$. The function $u$ is increasing because
if $a<b$ then,  as $(a,0)\le (b,0)$,  we have $(u(a),0)\le (u(b),0)$
and hence $u(a)\le u(b)$. The possibility $u(a)=u(b)$ is ruled out
for otherwise $(u(a),0)= (u(b),0) $ in contradiction with the
injectivity of $G$. Analogously, $v$ is increasing.

Let us prove that $G((x^+,x^-))=(u(x^+),v(x^-))$. Indeed,
$(x^+,0)\le (x^+,x^-)$ which implies after application of $G$,
$(u(x^+),0)\le G((x^+,x^-))$. Analogously, from $(0,x^-)\le
(x^+,x^-)$ we get $(0,v(x^-))\le G((x^+,x^-))$.  Therefore, we
deduce that $G((x^+,x^-))\in J^+((u(x^+),0))\cap
J^+((0,v(x^-)))=J^+((u(x^+),v(x^-))$ which means that
$(u(x^+),v(x^-))\le G((x^+,x^-))$. Now consider the point $r'=
(u(x^+),v(x^-))$, since $F$ is surjective, $G$ is surjective thus
there is $r\in \mathbb{R}^2$ such that $G(r)=r'$. Moreover, $r'\le
G((x^+,x^-))$ thus, as $G^{-1}$ is a causal automorphism, $r \le
(x^+,x^-)$. However, $(u(x^+),0)\le (u(x^+),v(x^-))=r'$ and
$(0,v(x^-))\le (u(x^+),v(x^-))=r'$ thus applying $G^{-1}$ we get
$(x^+,0)\le r$ and $(0,x^-)\le r$ from which it follows
$(x^{+},x^-)\le r$. We conclude that $r=(x^{+},x^-)$ and hence that
$G((x^+,x^-))=(u(x^+),v(x^-))$.

From this expression since $G$ is surjective it follows that both
$u$ and $v$ are surjective, and since $u$ and $v$ are surjective and
increasing they are continuous. Moreover, being bijective and
continuous they are homeomorphisms of the real line (invariance of
domain theorem). As a final step define $f= u+F(O)^{+}$ and $g
=v+F(O)^{-}$.

\end{proof}

Let $B$ denote the group of all causal automorphisms of the
Minkowski plane and let $N$ be the subgroup of causal automorphisms
made of maps $(x^+,x^-) \to (f(x^+),g(x^-))$ where
$f,g:\mathbb{R}\to \mathbb{R}$ are increasing homeomorphisms of the
real line. Denote with $S_2$ the symmetric group of permutations of
$\{x^+,x^-\}$ which is also a subgroup of all causal automorphisms.
It can be easily checked that $N$ is a normal subgroup of $B$.
Indeed, since we have proved that $B=N S_2$ we have only to show
that for $n\in N$ and $p\in S_2$, with $p$ the non trivial
transposition, we have $p n p \in N$. Indeed, let $n$ be the map
$(x^+,x^-) \to (f(x^+),g(x^-))$, then $n p$ is the map $(x^+,x^-)
\to (f(x^-),g(x^+))$ and $p n p$ is the map $(x^+,x^-) \to
(g(x^+),f(x^-))$ which belongs to $N$. As a consequence

\begin{corollary}
The group of causal automorphism of the Minkowski plane is
\[Hom_{\le}(\mathbb{R})^2 \rtimes S_2\] namely the semidirect product
between $Hom_{\le}(\mathbb{R})^2$ and $S_2$ where
$Hom_{\le}(\mathbb{R})$ is the group of order preserving
homeomorphism of the real line.
\end{corollary}

\section{Conclusions}
We have given a simple characterization of causal automorphisms of
Minkowski 1+1 spacetime. Theorem \ref{theo:main-result} can be
generalized, with very minor changes, to the case of a poset $P$
which is the cartesian $k$-product, $k\in\mathbb N$, of a totally
ordered poset $X$. Nevertheless, the spacetime interpretation of the
product holds only in the case $k=2$ considered here.


\section*{Acknowledgments}

E.M. thanks the Department of Mathematics of the Universidade do
Minho for kind hospitality. E.M. is partially supported by GNFM of
INDAM and by FQXi. Both authors acknowledge the financial support
provided by the Research Centre of Mathematics of the University of
Minho (Portugal) through the ``Funda\c{c}\~ao para a Ci\^encia e a
Tecnologia'' (FCT) Pluriannual Funding Program.

\end{document}